\documentclass[nobibnotes,aps,pre,showpacs,
twocolumn]{revtex4}
\usepackage{graphics}
\usepackage{graphicx}
\usepackage{color}
\usepackage{amsmath}
\usepackage{amssymb}
\usepackage{bm}
\usepackage{float}
\usepackage{epsfig}
\usepackage{epstopdf}
\usepackage{color,graphics}
\usepackage[toc,page]{appendix}

\newcommand\be{\begin{equation}}
\newcommand\e{\end{equation}}
\newcommand\ba{\begin{eqnarray}}
\newcommand\ay{\end{eqnarray}}

\begin{document}

\title{Nonlinear ion-acoustic solitons in a magnetized quantum plasma with
arbitrary degeneracy of electrons}
\author{Fernando Haas}
\affiliation{Physics Institute, Federal University of Rio Grande do Sul, CEP 91501-970,
Av. Bento Gon\c{c}alves 9500, Porto Alegre, RS, Brazil}
\author{Shahzad Mahmood}
\affiliation{Theoretical Physics Division (TPD), PINSTECH, P. O. Nilore Islamabad 44000,
Pakistan}

\begin{abstract}
Nonlinear ion-acoustic waves are analyzed in a non-relativistic magnetized
quantum plasma with arbitrary degeneracy of electrons. Quantum statistics is
taken into account by means of the equation of state for ideal fermions at
arbitrary temperature. Quantum diffraction is described by a modified Bohm
potential consistent with finite temperature quantum kinetic theory in the
long wavelength limit. The dispersion relation of the obliquely propagating
electrostatic waves in magnetized quantum plasma with arbitrary degeneracy
of electrons is obtained. Using the reductive perturbation method, the
corresponding Zakharov-Kuznetsov equation is derived, describing obliquely
propagating two-dimensional ion-acoustic solitons in a magnetized quantum
plasma with degenerate electrons having arbitrary electron temperature. It
is found that in the dilute plasma case only electrostatic potential hump
structures are possible, while in dense quantum plasma in principle both hump and dip
soliton structures are obtainable, depending on the electron plasma density
and its temperature. The results are validated by comparison with the
quantum hydrodynamic model including electron inertia and magnetization
effects. Suitable physical parameters for observations are identified. 
\end{abstract}

\maketitle

\section{Introduction}

The ion-acoustic wave, which is the fundamental low frequency mode of plasma
physics, is a prime focus of many current studies of localized electrostatic
disturbances in laboratory, space and astrophysical plasmas. The study of
ion-acoustic waves has also gained its importance in quantum plasmas to
understand electrostatic wave propagation in microscopic scales. During the
last decade, there has been a renewed interest to study collective wave
phenomenon in quantum plasma, motivated by applications in semiconductors \cite{r1},
high intensity laser-plasma experiments \cite{r3,r4,gregori} and high density astrophysical plasmas 
such as in the interior of massive planets and white dwarfs, neutron stars or magnetars \cite{r5,r6,r7}. 
The quantum or degeneracy effects appears in plasmas when the de Broglie wavelength
associated with the charged carriers becomes of the order of the
inter-particle distances. The quantum effects in plasmas are more frequently due
to electrons, which are lighter than ions, and it includes both Pauli's
exclusion principle (for half spin particles) and Heisenberg's uncertainty
principle, due to wave like nature of the particles.

Quantum ion-acoustic waves in unmagnetized dense plasma have been
investigated using quantum hydrodynamic models \cite{r8}. In quantum
hydrodynamics, the momentum equation for degenerate electrons contains a
pressure term compatible with a Fermi-Dirac distribution function, while the
Bohm potential term is included to account for quantum diffraction \cite{r9,r10,r12}. 
Later on the quantum hydrodynamics model for plasmas was
extended to include magnetic fields, with the associated quantum
magnetohydrodynamics theory developed and discussed in connection to
astrophysical dense plasmas \cite{r14}. Quantum Trivelpiece-Gould modes in a
dense magnetized quantum plasma were derived \cite{Tercas}. The exchange
effects on low frequency excitations in plasma have been discussed \cite{zama1}, 
using a modified Vlasov equation incorporating the exchange
interaction \cite{zama2, zama3}.

The Zakharov-Kuznetsov (ZK) equation was derived in 1974, to study nonlinear
propagation of ion-acoustic waves in magnetized plasmas \cite{r20}-\cite{Laedke}. 
The ZK equation is a multi-dimensional extension of the well-known
Korteweg-de Vries equation for studying solitons (or single pulse
structures). In the degenerate magnetized plasma case, the cold
Fermi electron gas assumption has been applied in the derivation of the
appropriate ZK equation \cite{mos1, mos2, kha}, restricted to the fully
degenerate case of negligible thermodynamic temperature in comparison to the
Fermi temperature.

Linear ion-acoustic and electron Langmuir waves in a plasma with arbitrary
degeneracy of electrons were studied using quantum kinetic theory \cite{r28}. 
The nonlinear theory of the isothermal ion-acoustic waves in degenerate
unmagnetized electron plasmas was investigated \cite{r29}. The ranges of the
phase velocities of the periodic ion-acoustic waves and the soliton speed
were determined in degenerate plasma, but ignoring quantum diffraction
effects. Also nonlinear Langmuir waves in a dense plasma with arbitrary
degeneracy of electrons in the absence as well as in the presence of quantum
diffraction effects in the model have been studied \cite{r31}. Eliasson and
Shukla \cite{r33} derived certain nonlinear quantum electron fluid equation by
taking into account the moments of the Wigner equation and using the
Fermi-Dirac distribution function for electrons with arbitrary temperature.
The relativistic description of localized wavepackets in electrostatic
plasma \cite{mckerr} as well as the associated ZK equation for dense
relativistic plasma \cite{Behery} were obtained, in the limit of
negligible thermodynamic temperature. Recently, the hydrodynamic equations
for ion-acoustic excitations in electrostatic quantum plasma with arbitrary
degeneracy were put forward \cite{pre}. The purpose of the present
communication is to achieve a notable generalization of this work, obtaining
the corresponding fluid theory for quantum magnetized ion-acoustic waves
(MIAWs), both in the linear and nonlinear realms. The extension has a
definite interest since magnetized degenerate plasmas are ubiquitous in
astrophysics as well as in laboratory \cite{Fortov}. Specifically, it is of
fundamental interest to access the nonlinear aspects of quantum MIAWs, which
is a more accessible trend using hydrodynamic methods.

The manuscript is organized in the following way. In Section II, the set of
dynamic equations or studying ion-acoustic waves in magnetized quantum
plasmas with arbitrary degeneracy of electrons is presented. In Section III,
the dispersion relation of the obliquely propagating electrostatic linear
waves in magnetized quantum plasma with arbitrary degeneracy of electrons is
obtained. The limiting cases of waves parallel or perpendicular to the the
magnetic field are discussed, as well as the strongly magnetized ions limit.
Section IV describes the modifications of the linear dispersion relation due
to the inclusion of electron inertia and magnetization effects. Section V shows
that the fluid theory is the limit case of quantum kinetic theory in the
long wavelength limit, as it should be. In Section VI, using reductive
perturbation methods, the ZK equation for two dimensional propagation of
nonlinear ion-acoustic waves is derived for a magnetized degenerate
electrons plasma with arbitrary temperature. The soliton solution is
presented. Section VII illustrates the results, using suitable plasma parameters
for observations, within the applicability range of the model. Section VIII contains the
summary of the conclusions. Finally, Appendix A has a more detailed derivation of the static
electronic response from quantum kinetic theory, necessary in Section VI.

\section{Dynamic equations for magnetized quantum fluids}

Consider a quantum electron-ion plasma with arbitrary degeneracy of
electrons, embedded in an external magnetic field $\mathbf{B}_{0}=B_{0}\hat{x}$ 
directed along the x-axis. In principle, the electrostatic wave is assumed to propagate obliquely to the external
magnetic field in the $xy$-plane i.e., $\mathbf{\nabla} =(\partial_x,\partial_y,0)$. In order to study the quantum MIAWs the ions are
taken to be inertial, while electrons are assumed to be inertialess. 
The set of dynamic equations for MIAWs in a quantum plasma with arbitrary
degeneracy of electrons is described as follows.

The ion continuity equation is given by 
\begin{equation}
\frac{\partial n_{i}}{\partial t}+\frac{\partial }{\partial x}(n_{i}u_{ix})+
\frac{\partial }{\partial y}(n_{i}u_{iy})=0 \,,  \label{e1}
\end{equation}
while the ion momentum equations in component form are 
\begin{eqnarray}
\frac{\partial u_{ix}}{\partial t}+\left( u_{ix}\frac{\partial }{\partial x}
+u_{iy}\frac{\partial }{\partial y}\right) u_{ix}=-\frac{e}{m_{i}}\frac{\partial \phi }{\partial x} \,,  \label{e2}
\\
\frac{\partial u_{iy}}{\partial t}+\left( u_{ix}\frac{\partial }{\partial x}
+u_{iy}\frac{\partial }{\partial y}\right) u_{iy}=-\frac{e}{m_{i}}\frac{\partial \phi }{\partial y}+\omega _{ci}u_{iz} \,,  \label{e3}
\\
\frac{\partial u_{iz}}{\partial t}+\left( u_{ix}\frac{\partial }{\partial x}
+u_{iy}\frac{\partial }{\partial y}\right) u_{iz}=-\omega _{ci}u_{iy} \,.
\label{e4}
\end{eqnarray}
The momentum equation for the inertialess quantum electron fluid is 
\begin{equation}
0 = - \frac{\nabla p}{n_{e}} + e\nabla\phi + \frac{\alpha \hbar ^{2}}{6m_{e}}
\nabla \left[\frac{1}{\sqrt{n_{e}}}\left(\frac{\partial^{2}}{\partial x^{2}}
+ \frac{\partial^{2}}{\partial y^{2}}\right)\sqrt{n_{e}}\right] \,.
\label{e5}
\end{equation}
The Poisson equation is written as 
\begin{equation}
\left( \frac{\partial ^{2}}{\partial x^{2}}+\frac{\partial ^{2}}{\partial
y^{2}}\right) \phi =\frac{e}{\varepsilon _{0}}(n_{e}-n_{i}),  \label{e6}
\end{equation}
where $\phi$ is the electrostatic potential. The ion fluid density and
velocity are represented by $n_{i}$ and ${\bf u}_{i}=(u_{ix},u_{iy},u_{iz})$ respectively, while $n_{e}$ is the electron
fluid density. Also, $m_{e}$ and $m_{i}$ are the electron and ion masses, $-e $ is the electronic charge, $\varepsilon _{0}$ is the vacuum
permittivity, $\hbar $ is the reduced Planck's constant and $\omega_{ci}=eB_{0}/m_{i}$ is the ion cyclotron frequency. In equilibrium, we
have $n_{e0}=n_{i0}\equiv n_{0}$. The electron's fluid pressure $p=p(n_{e})$
is specified by a barotropic equation of state which is given below. The
last term on the right hand side of the momentum equation (\ref{e5}) for
electrons is the quantum force, which arises from the Bohm potential, giving
rise to quantum diffraction or tunneling effects due to the wave like nature
of the charged particles. The dimensionless quantity $\alpha $ is selected
in order to fit the kinetic linear dispersion relation in the long
wavelength limit, in a Fermi-Dirac equilibrium, as shown in the
continuation. Quantum effects on ions are ignored in view of their large
mass in comparison to electrons. In addition, temperature effects on ions
are disregarded. Finally, to avoid too much complexity and to focus on the
interplay between degeneracy and quantum recoil, exchange effects are also
ignored.

The equation of state can be obtained from the moments of a local
Fermi-Dirac distribution function \cite{pre, r27} of an ideal Fermi gas and
reads 
\begin{equation}
p = \frac{n_{e}}{\beta }\,\frac{{\rm Li}_{5/2}(-e^{\beta \mu })}{{\rm %
Li}_{3/2}(-e^{\beta \mu })} \,,  \label{e12}
\end{equation}
where $\beta = (\kappa_B T)^{-1}$, $\kappa_B$ is the Boltzmann constant, $T$
is the temperature and $\mu$ is the chemical potential, satisfying 
\begin{equation}
n_{e}=n_{0}\,\frac{{\rm Li}_{3/2}(-e^{\beta \mu })}{{\rm Li}
_{3/2}(-e^{\beta \mu _{0}})} \,.  \label{e121}
\end{equation}
The equilibrium chemical potential $\mu_0$ is related to the
equilibrium density $n_0$ through 
\begin{equation}  \label{muu}
- \,\frac{n_0}{{\rm Li}_{3/2}(-e^{\beta \mu_0 })}\left(\frac{\beta m_{e}}{%
2\pi} \right) ^{3/2} = 2\left( \frac{m_{e}}{2\pi \hbar }\right) ^{3} = {\cal A} \,,
\end{equation}
where the quantity ${\cal A}$ was defined for later convenience.

Equations (\ref{e12}) and (\ref{e121}) contain the polylogarithm function 
${\rm Li}_{\nu}(- z)$ with index $\nu $, which for $\nu > 0$ can be
defined \cite{Lewin} as 
\begin{equation}
{\rm Li}_{\nu }(-z )=-\frac{1}{\Gamma (\nu )}\int_{0}^{\infty }
\frac{s^{\nu -1}}{1 + e^{s}/z}ds \,, \quad \nu > 0  \label{e9}
\end{equation}
where $\Gamma(\nu)$ is the gamma function. For $\nu < 0$ one applies 
\begin{equation}  \label{shi}
{\rm Li}_{\nu}(-z) = \left(z \frac{\partial}{\partial z}\right){\rm Li}
_{\nu+1}(-z)
\end{equation}
as many times as necessary, where $\nu+1>0$.

\begin{figure}[!hbt]
\begin{center}
\includegraphics[width=8.0cm,height=6.0cm]{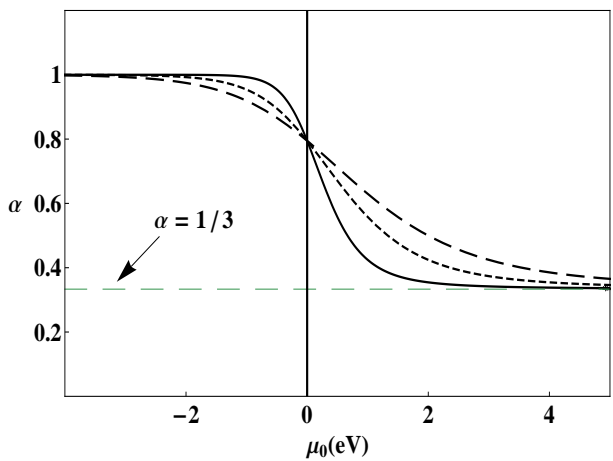}
\end{center}
\caption{Coefficient $\alpha$ from Eq. (\ref{e61}), as a function of the chemical potential $\mu_0$.
Solid line: $T = 4000 \, {\rm K}$. Doted line: $T = 6000 \, {\rm K}$. Dashed line: $T = 8000 \, {\rm K}$.}
\label{fig1}
\end{figure}

The numerical coefficient $\alpha $ appearing in the Bohm potential term in
Eq. (\ref{e5}) has been derived from finite-temperature quantum kinetic
theory for low-frequency electrostatic excitations in \cite{pre}, 
\begin{equation}
\alpha =\frac{{\rm Li}_{3/2}(-e^{\beta \mu _{0}})\, {\rm Li}
_{-1/2}(-e^{\beta \mu _{0}})}{[{\rm Li}_{1/2}(-e^{\beta \mu _{0}})]^{2}},
\label{e61}
\end{equation}
expressed as a function of the equilibrium fugacity $z = \exp (\beta \mu
_{0})$. The treatment of \cite{pre} considered non-magnetized plasmas, but
in Section \ref{kin} it is proved that Eq. (\ref{e61}) applies to MIAWs too.
As discussed in \cite{pre}, in the classical limit ($z \ll 1$) one has 
$\alpha \approx 1$, while in the full degenerate limit ($z \gg 1$) one has $%
\alpha \approx 1/3$. The same behavior is seen for $\alpha$ as a function of
the chemical potential, as depicted in Fig. \ref{fig1}, showing a transition
zone from classical to dense regimes.

It happens \cite{Bengt} that the finite-temperature quantum hydrodynamic equations using $\alpha$ from Eq. (\ref{e61}) are 
consistent with the results from orbital free density functional theory \cite{mur1, mur2}. 

\section{Linear waves}
\label{lin}

In order to find the dispersion relation for electrostatic wave in a
magnetized quantum plasma with arbitrary degeneracy of electrons, we
linearize the system of equations (\ref{e1})-(\ref{e6}) by considering 
\begin{eqnarray}
n_{i} &=&n_{0}+n_{i1} \,, \quad n_{e}=n_{0}+n_{e1} \,, \quad
u_{ix}=u_{ix1} \,, \nonumber \\
u_{iy} &=&u_{iy1} \,, \quad u_{iz}=u_{iz1} \,, \quad \phi =\phi _{1} \,,
\end{eqnarray}
inducing a correction $\mu =\mu _{0}+\mu _{1}$, where the subscript $1$
denotes the first order quantities. In particular, using the expansion of
the polylogarithm function  
to first order, i.e., 
\begin{equation}
{\rm Li}_{\nu }(-e^{\beta (\mu _{0}+\mu _{1})})={\rm Li}_{\nu
}(-e^{\beta \mu _{0}})+\beta \mu _{1}{\rm Li}_{\nu -1}(-e^{\beta \mu
_{0}}) \,,
\end{equation}
and considering plane wave perturbations $\sim \exp [i(k_{x}x+k_{y}y-\omega
t)]$, the result is 
\begin{equation}  \label{dis}
1 + \chi_{i}(\omega,{\bf k}) + \chi_{e}(0,{\bf k}) = 0 \,,
\end{equation}
where the ionic and electronic susceptibilities are respectively given by 
\begin{eqnarray}
\chi_{i}(\omega,{\bf k}) &=& - \, \frac{\omega_{pi}^2 (\omega^2 -
\omega_{ci}^2 \cos^{2}\theta)}{\omega^2 (\omega^2 - \omega_{ci}^2)} \,,
\label{ci} \\
\chi_{e}(0,{\bf k}) &=& \omega_{pe}^2 \left[\frac{1}{m_e}\left(\frac{dp}{dn_e}\right)_{0}k^2 + \frac{\alpha \hbar^2 k^4}{12 m_{e}^2}\right]^{-1}
\!\!\!,  \label{ce}
\end{eqnarray}
where $\omega_{pj}^2 = n_{0}e^2/(m_j \varepsilon_0)$ for $j = i, e$ and 
${\bf k} = k \,(\cos\theta,\sin\theta,0)$. Due to the neglect of electrons
inertia, only the static electronic susceptibility $\chi_{e}(0,{\bf k})$
is necessary. There is no loss of generality in assuming waves in the $xy$
plane, due to the cylindrical geometry around the $x-$axis.

The dispersion relation (\ref{dis}) develops as a quadratic equation for $%
\omega^2$ whose solution is 
\begin{eqnarray}  \label{w}
\omega^2 = \frac{1}{2}\Bigl[\omega_{0}^2 &+& \omega_{ci}^2 \\
&\pm& \Bigl((\omega_{0}^2 + \omega_{ci}^2)^2 - 4 \omega_{0}^2 \omega_{ci}^2
\cos^{2}\theta\Bigr)^{1/2}\Bigr] \,,  \nonumber
\end{eqnarray}
where $\omega_0$ was already obtained \cite{pre} in the case of unmagnetized
quantum ion-acoustic waves, 
\begin{equation}  \label{w0}
\omega_{0}^2 = \frac{c_{s}^2 k^2 \left[1 + H^2 (k\lambda_D)^2/4\right]}{1 +
(k\lambda_D)^2 + H^2 (k\lambda_D)^4/4} \,.
\end{equation}
In Eq. (\ref{w0}) one has the ion-acoustic speed $c_s$ which follows from 

\begin{equation}
c_{s}^2 = \frac{1}{m_i}\left(\frac{dp}{dn_e}\right)_0 = \frac{\kappa_B T}{m_i}
\,\frac{{\rm Li}_{3/2}(-e^{\beta \mu _{0}})} {{\rm Li}
_{1/2}(-e^{\beta \mu _{0}})} \,,  \label{cs}
\end{equation}
the generalized electronic screening length $\lambda_D$ from 
\begin{equation}  \label{ld}
\lambda_{D}^2 = \frac{c_{s}^2}{\omega_{pi}^2} = \frac{\kappa_B T}{m_e
\omega_{pe}^2}\,\frac{{\rm Li}_{3/2}(-e^{\beta \mu _{0}})} {{\rm Li}
_{1/2}(-e^{\beta \mu _{0}})} \,,
\end{equation}
as well as the quantum diffraction parameter $H$ specified by 
\begin{equation}
H=\frac{\beta \hbar \omega _{pe}}{\sqrt{3}}\left( \frac{{\rm Li}
_{-1/2}(-e^{\beta \mu _{0}})}{{\rm Li}_{3/2}(-e^{\beta \mu _{0}})}\right)
^{1/2} \,.  \label{e361}
\end{equation}

In the dilute plasma limit $e^{\beta \mu _{0}} \ll 1$, implying ${\rm 
Li}_{\nu}(- e^{\beta\mu_0}) \approx - e^{\beta\mu_0}$, one has $c_s \approx 
\sqrt{\kappa_B T/m_i}, \, \lambda_D = \sqrt{\kappa_B T/(m_e \omega_{pe}^2)}$, which respectively are the 
more traditional ion-acoustic speed and Debye length, and $H \approx \beta\hbar\omega_{pe}/\sqrt{3}$. On the
other hand, in the fully de\-ge\-ne\-ra\-te case $e^{\beta \mu _{0}} \gg 1$, using 
${\rm Li}_{\nu}(- e^{\beta\mu_0}) \approx - (\beta\mu_0)^{\nu}/\Gamma(\nu
+ 1)$ one has $\mu_0 \approx E_F = \hbar^2 (3\pi^2 n_0)^{2/3}/(2 m_e)$, which is the
Fermi energy, and $c_s \approx \sqrt{(2/3)E_{F}/m_i}, \, \lambda_D = \sqrt{2
E_F/(3 m_e \omega_{pe}^2)}$, which are respectively the quantum ion-acoustic speed and 
the Thomas-Fermi screening length, and $H \approx (1/2)\hbar\omega_{pe}/E_F$. The dispersion relation (\ref{w}) is
formally the same as for classical magnetized plasma \cite{Stringer, Witt},
provided the fully quantum ion-acoustic frequency $\omega_0$ is replaced by
its purely classical counterpart.

As apparent from the dispersion relation (\ref{dis}), ions are responsible
for providing inertia effects, while electrons are responsible for kinetic
energy (arising from the standard, thermodynamic temperature and/or Fermi pressure) 
and quantum diffraction is represented by the parameter 
$H$. It is convenient to rewrite Eq. (\ref{e361}) using Eq. (\ref{muu}),
yielding 
\begin{equation}  \label{hh}
H^2 = - \frac{2 \alpha_F}{3}\,\sqrt{\frac{2 \beta m_e c^2}{\pi}}\,{\rm Li}
_{-1/2}(-e^{\beta\mu_0}) \,,
\end{equation}
where $\alpha_F = e^2/(4\pi\varepsilon_0\hbar c) \approx 1/137$ is the fine
structure constant. Obviously the theory is non-relativistic, in spite of
the appearance of the rest energy $m_e c^2$ in Eq. (\ref{hh}).

For a fixed temperature, $H$ is a simple function of the fugacity $z =
\exp(\beta\mu_0)$, as shown in Fig. \ref{fig2} below. It is seen that the
pure wave like quantum effects are enhanced for larger densities (and
fugacities) up to $z \approx 3.03$, while for larger degeneracy the quantum
statistical effects prevail showing that the quantum force becomes less
effective in denser systems, in view of Pauli's exclusion principle.
Therefore for dilute systems $H$ increases with the density and decreases
with temperature, while for fully degenerate systems the leading order behavior shows a decreasing of $H$ for 
an increasing density. 

\begin{figure}[!hbt]
\begin{center}
\includegraphics[width=8.0cm,height=6.0cm]{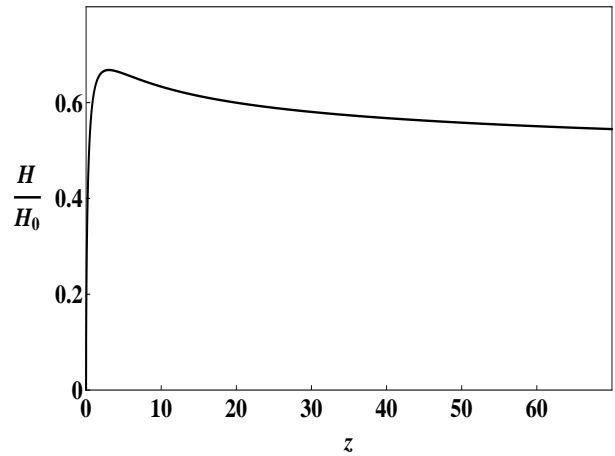}
\end{center}
\caption{Quantum diffraction parameter defined in Eq. (\ref{hh}), as
a function of the equilibrium fugacity $z = e^{\beta\mu_0}$, and
normalized to $H_0 = [(2\alpha_F/3) (2\beta m_e c^2/\pi)^{1/2}]^{1/2}$.}
\label{fig2}
\end{figure}

The positive sign in Eq. (\ref{w}) corresponds to fast electrostatic waves,
and a negative sign corresponds to slow electrostatic waves in a magnetized
plasma.  The effect of degeneracy for arbitrary angle $\theta$ is not entirely straightforward 
to identify, due to the somehow involved expression (\ref{w}). However, in practical applications it is likely to 
have $\omega_0 \gg \omega_{ci}$, so that the fast mode becomes $\omega^2 \approx \omega_{0}^2 + \omega_{ci}^2 \sin^{2}\theta$ while the slow mode becomes $\omega^2 \approx \omega_{ci}^2 \cos^{2}\theta$. Since quantum effects are present only on $\omega_{0}^2 \gg \omega_{ci}^2$, it happens that the fast wave has an angular dependence appearing as a correction, while the slow wave is strongly angle dependent, but not so influenced by quantum effects. As an example, consider hydrogen plasma with an ambient magnetic field $B_0 = 10^3 \,{\rm T}$, yielding the ion cyclotron frequency $\omega_{ci} = 9.58 \times 10^{10} \,{\rm rad/s}$.  For solid-density plasma with $T = 10^6 \,{\rm K}, \, n_0 = 5 \times 10^{30}\, {\rm m}^{-3}$  and a wave-number $k = 2\pi \times 10^{9} \, {\rm m}^{-1}$, one has $\omega_0 = 6.50 \times 10^{14} \, {\rm rad/s} \gg \omega_{ci}$. These parameters are in accordance with the more detailed validity conditions discussed in Section VII. 

In addition, some significant limiting cases are described below.

\subsection{Wave propagation parallel to the magnetic field}
\label{para}

Considering $\theta = 0$ in Eq. (\ref{w}), one has either $\omega =
\omega_{ci}$ (the ion cyclotron frequency) or $\omega = \omega_{0}$. The later
has already been discussed in \cite{pre}. In the limit $k\lambda_D \ll 1$
one has $\omega_0 \approx c_s k$, where the ion-acoustic speed contains only
quantum degeneracy effects as seen from Eq. (\ref{cs}).

\subsection{Wave propagation perpendicular to the magnetic field}

\label{perp}

Considering $\theta = \pi/2$ in Eq. (\ref{w}), one has a vanishing solution (%
$\omega^2 = 0$) as well as a quantum modified electrostatic ion cyclotron
wave given by $\omega^2 = \omega_{ci}^2 + \omega_{0}^2$.

\subsection{Strongly magnetized ions}

For completeness we consider the strongly magnetized ions case. If $\omega_{ci} \gg \omega_0$ one has the fast mode 
\begin{equation}
\omega^2 = \omega_{ci}^2\left[1 + \frac{\omega_{0}^2}{\omega_{ci}^2}
\sin^{2}\theta + \frac{\omega_{0}^4}{4\,\omega_{ci}^4}\sin^{2}(2\theta) + 
\mathcal{O}\left(\left(\frac{\omega_{0}}{\omega_{ci}}\right)^6\right)\right]
\,,
\end{equation}
and the slow mode 
\begin{equation}
\omega^2 = \omega_{0}^2 \cos^{2}\theta\left[1- \frac{\omega_{0}^2}
{\omega_{ci}^2}\sin^{2}\theta + \mathcal{O}\left(\left(\frac{\omega_{0}}
{\omega_{ci}}\right)^4\right)\right] \,.
\end{equation}

\section{Electron inertia and magnetization effects}
\label{ine}

In the purely classical case, the conditions of applicability of the model
are well-known \cite{r21, Laedke}. In the quantum case, it is interesting to
include electron inertia and magnetization effects, to measure the
limitations of Eq. (\ref{dis}). In this context one adds to Eqs. 
(\ref{e1})-(\ref{e4}) and (\ref{e6}) the electron continuity equation 
\begin{equation}
\frac{\partial n_{e}}{\partial t}+\frac{\partial }{\partial x}(n_{e}u_{ex})+
\frac{\partial }{\partial y}(n_{e}u_{ey})=0 \,,
\end{equation}
and replace Eq. (\ref{e5}) by 
\begin{eqnarray}
m_e \Bigl(\frac{\partial{\bf u}_e}{\partial t} &+& {\bf u}_e \cdot
\nabla{\bf u_e}\Bigr) = - \frac{\nabla p}{n_{e}} - e\Bigl(-\nabla\phi + 
{\bf u}_e \times {\bf B}_0\Bigr)  \nonumber \\
&+& \frac{\alpha \hbar^{2}}{6m_{e}}\nabla \Bigl[\frac{1}{\sqrt{n_{e}}}
\Bigl(\frac{\partial^{2}}{\partial x^{2}} + \frac{\partial^{2}}{\partial
y^{2}}\Bigr)\sqrt{n_{e}}\Bigr] \,,
\end{eqnarray}
where ${\bf u}_e = (u_{ex},u_{ey},u_{ez})$ is the electron fluid
velocity. Proceeding as in Section \ref{lin} and also supposing linear
perturbations where ${\bf u}_e = {\bf u}_{e1}$, the result is 
\begin{equation}  \label{diss}
1 + \chi_{i}(\omega,{\bf k}) + \chi_{e}(\omega,{\bf k}) = 0 \,,
\end{equation}
where the ionic susceptibility is still given by Eq. (\ref{ci}) and 
\begin{equation}
\chi_{e}(\omega,{\bf k}) = - \frac{\omega_{pe}^2 (\omega^2 -
\omega_{ce}^2 \cos^{2}\theta)}{\omega^4 - (k^2 v_{T}^{2}(k) +
\omega_{ce}^2)\omega^2 + k^2 v_{T}^{2}(k) \omega_{ce}^2 \cos^{2}\theta} \,,
\label{cce}
\end{equation}
where $\omega_{ce} = eB_0/m_e$ is the electron cyclotron frequency and 
\begin{equation}  \label{vt}
v_{T}^2(k) = \frac{1}{m_e}\left(\frac{dp}{dn_e}\right)_{0} + \frac{\alpha
\hbar^2 k^2}{12 m_{e}^2} \,.
\end{equation}
It is not the purpose of this work to develop the full consequences of the
dispersion relation (\ref{diss}), but it is useful to observe that in the
formal limit $\omega \rightarrow 0$ the electron response (\ref{cce})
regains the static electronic response $\chi_{e}(0,{\bf k})$ given by Eq.
(\ref{ce}). Moreover, by inspection of Eq. (\ref{cce}) it is found that such
a limit is attended for a warm electron fluid, where $k^2 v_{T}^{2}(k) \gg
\omega_{ce}^2$ so that the electrons magnetization could be disregarded, and 
$k^2 v_{T}^{2}(k) \gg \omega^2$, which is attainable for low frequency
excitations.  In addition, notice that $v_{T}(k)$ from Eq.
(\ref{vt}) depends not only on pressure but also on quantum diffraction
effects. On the other hand, ions are assumed to
be cold and non-quantum enough.

\section{Comparison to kinetic theory}
\label{kin}

The results from hydrodynamics should agree with kinetic theory in the long
wavelength limit. Therefore it is necessary to compare the ionic and
electronic responses found from kinetic theory, to the susceptibilities
shown in Eqs. (\ref{ci}) and (\ref{ce}). Since ions are safely assumed as
classical in most cases, their particle distribution function $f_i = f_{i}({\bf r}, 
{\bf v}, t)$ satisfy Vlasov's equation, which presently is 
\begin{equation}
\left[\frac{\partial}{\partial t} + {\bf v}\cdot\nabla + \frac{e}{m_i}(-
\nabla\phi + {\bf v}\times{\bf B}_0)\cdot\frac{\partial}{\partial{\bf v}}\right]f_i= 0 \,.
\end{equation}

On the other hand, the quantum nature of electrons deserves the use of the
quantum Vlasov equation satisfied by the electronic Wigner
quasi-distribution $f_e = f_{e}({\bf r}, {\bf v}, t)$, 
\begin{eqnarray}
\frac{\partial f_e}{\partial t} &+& {\bf v}\cdot\nabla f_e - \frac{i e}
{\hbar}\left(\frac{m_e}{2\pi\hbar}\right)^3 \times  \label{qv} \\
&\times& \int d{\bf s} \,d{\bf v}^{\prime }\exp\left(\frac{i m_e ({\bf v}^{\prime }- {\bf v})\cdot{\bf s}}{\hbar}\right) \times 
\nonumber \\
&\times& \left[\phi\left({\bf r}+\frac{{\bf s}}{2}, t\right) -
\phi\left({\bf r}-\frac{{\bf s}}{2}, t\right)\right]f_{e}({\bf r},
{\bf v}^{\prime },t) = 0 \,.  \nonumber
\end{eqnarray}
All integrals run from $-\infty$ to $\infty$, unless otherwise stated.
Moreover, under the same assumption as before, namely large electron thermal
and quantum (statistical and diffraction) effects, the magnetic force on
electrons was omitted in Eq. (\ref{qv}).

The scalar potential is self-consistently determined by Poisson's equation, 
\begin{equation}
\left( \frac{\partial ^{2}}{\partial x^{2}}+\frac{\partial ^{2}}{\partial
y^{2}}\right)\phi = \frac{e}{\varepsilon_0}\left(\int\! f_e \,d{\bf v} -
\int\! f_i \,d{\bf v}\right) \,,
\end{equation}
where we are taking spatial variations in the $xy$-plane only.

Proceeding as in Section \ref{lin}, assuming plane wave perturbations $\sim 
\exp[i(k_x x + k_y y - \omega t)]$ around isotropic in velocities
equilibria, the dispersion relation $1 + \chi_{i}(\omega,{\bf k}) +
\chi_{e}(\omega,{\bf k}) = 0$ is easily derived. Disregarding the
negligibly small Landau damping of MIAWs in the case of cold ions, it is found \cite%
{Stepanov, Akhiezer} that the ionic susceptibility from kinetic theory
coincides with the fluid expression (\ref{ci}). On the other hand, for low
frequency waves, the static limit $\chi_{e}(0,{\bf k})$ is sufficient for
electrons, reading 
\begin{equation}  \label{kce}
\chi_{e}(0,{\bf k}) = \frac{e^2}{\varepsilon_0 \hbar k^2} \int\!\frac{d%
{\bf v}}{{\bf k}\cdot{\bf v}}\!\left[F\left({\bf v}\!-\!\frac{%
\hbar{\bf k}}{2m_e}\right) - F\left({\bf v}\!+\!\frac{\hbar{\bf k}}{2m_e%
}\right)\right] ,
\end{equation}
where the principal value of the integral is understood if necessary and where the 
equilibrium electronic Wigner function is $f_e = F({\bf v})$.

Consider a Fermi-Dirac equilibrium,
\begin{equation}  \label{fd}
F({\bf v}) = \frac{\cal A}{1+e^{\beta (m_{e}v^{2}/2 -\mu_0 )}} \,, \quad v = |{\bf v}| \,,
\end{equation}
where the normalization constant ${\cal A}$ is given in Eq. (\ref{muu}), assuring
that $\int\!F({\bf v})\,d{\bf v} = n_0$. 

It turns out that the right-hand side of Eq. (\ref{kce}) can be evaluated as
a power series of the quantum recoil $q = \sqrt{\beta/(2 m_e)} \, \hbar k/2$, 
supposed to be a small quantity for long wavelengths and/or sufficiently
large electronic temperature: 
\begin{eqnarray}
\chi_{e}(0,{\bf k}) &=& \frac{\beta m_e \omega_{pe}^2}{\sqrt{\pi}\,{\rm Li}_{3/2}(-z)k^2} 
\Bigl[\Gamma\left(\frac{1}{2}\right){\rm Li}
_{1/2}(-z) +  \nonumber \\
&+& \Gamma\left(- \frac{1}{2}\right){\rm Li}_{- 1/2}(-z)\frac{q^2}{3} 
\nonumber \\
&+& \Gamma\left(- \frac{3}{2}\right){\rm Li}_{- 3/2}(-z)\frac{q^4}{5} +
\dots \Bigr]  \nonumber \\
&=& \frac{\beta m_e \omega_{pe}^2}{\sqrt{\pi}\,{\rm Li}_{3/2}(-z)k^2}
\sum_{j=0}^{\infty}\Gamma\left(\frac{1}{2}-j\right) \times  \nonumber \\
&\times& {\rm Li}_{1/2-j}(-z) \frac{q^{2j}}{2j+1} \,,  \label{sum}
\end{eqnarray}
where $z = e^{\beta\mu_0}$. The derivation
is detailed in the Appendix A. The expression (\ref{sum}) is exact, as long
as the series converges. Moreover, it coincides with the static limit of Eq.
(29) of \cite{MelMu}, where only the leading $\mathcal{O}(q^2)$ quantum
recoil correction was calculated.

For the sake of comparison, the hydrodynamic result from Eq. (\ref{ce}) can
be rewritten as 
\begin{eqnarray}
\chi_{e}(0,{\bf k}) &=& \frac{\beta m_e \omega_{pe}^2 {\rm Li}_{1/2}(-z)}
{\,{\rm Li}_{3/2}(-z)k^2}\left(1+\frac{2 q^2}{3}\frac{{\rm 
Li}_{-1/2}(-z)}{{\rm Li}_{1/2}(-z)}\right)^{-1}  \nonumber \\
&=& \frac{\beta m_e \omega_{pe}^2}{\,{\rm Li}_{3/2}(-z)k^2}\Bigl({\rm 
Li}_{1/2}(-z) - \frac{2 q^2}{3}{\rm Li}_{-1/2}(-z)  \nonumber \\
&+& \mathcal{O}(q^4)\Bigr) \,,
\end{eqnarray}
which coincides with Eq. (\ref{sum}) in the long wavelength limit, in view
of $\Gamma(1/2)=\sqrt{\pi},\,\, \Gamma(-1/2) = - 2\sqrt{\pi}$. This
completes the justification of $\alpha$ in Eq. (\ref{e61}) in the magnetized
case.

\section{Zakharov-Kuznetsov equation for arbitrary degeneracy}
\label{zkd}

In order to derive the ZK equation for obliquely propagating MIAWs in
arbitrary degenerate plasma, it is convenient to make use of normalized
quantities. The dispersion relation (\ref{w}) suggests the use of the
dimensionless variables $(\tilde{x},\tilde{y})=(x,y)/\lambda _{D},\,\,\,
\tilde{t}=\omega _{_{pi}}t$, $(\tilde{u}_{ix},\tilde{u}_{iy},\tilde{u}
_{iz})=(u_{ix},u_{iy},u_{iz})/c_{s}$ as well as $\tilde{\phi}=e\phi
/(m_{i}c_{s}^{2})$, $\tilde{n}_{j}=n_{j}/n_{0}$, where $j=e,i$. Equations 
(\ref{e1})-(\ref{e6}) are then written as 
\begin{eqnarray}
\frac{\partial \tilde{n}_{i}}{\partial \tilde{t}}+\frac{\partial }{\partial 
\tilde{x}}(\tilde{n}_{i}\tilde{u}_{ix})+\frac{\partial }{\partial \tilde{y}}(\tilde{n}_{i}\tilde{u}_{iy})=0 \,,  \label{e30}
\\
\frac{\partial \tilde{u}_{ix}}{\partial \tilde{t}}+\left( \tilde{u}_{ix}
\frac{\partial }{\partial \tilde{x}}+\tilde{u}_{iy}\frac{\partial }{\partial 
\tilde{y}}\right) \tilde{u}_{ix}=-\frac{\partial \tilde{\phi}}{\partial 
\tilde{x}} \,,  \label{e31}
\\
\frac{\partial \tilde{u}_{iy}}{\partial \tilde{t}}+\left( \tilde{u}_{ix}
\frac{\partial }{\partial \tilde{x}}+\tilde{u}_{iy}\frac{\partial }{\partial 
\tilde{y}}\right) \tilde{u}_{iy}=-\frac{\partial \tilde{\phi}}{\partial 
\tilde{y}}+\Omega \tilde{u}_{iz} \,,  \label{e32}
\\
\frac{\partial \tilde{u}_{iz}}{\partial \tilde{t}}+\left( \tilde{u}_{ix}
\frac{\partial }{\partial \tilde{x}}+\tilde{u}_{iy}\frac{\partial }{\partial 
\tilde{y}}\right) \tilde{u}_{iz}=-\Omega \tilde{u}_{iy} \,,  \label{e33}
\\
0=\tilde{\nabla}\tilde{\phi} -\frac{{\rm Li}_{1/2}(-e^{\beta \mu _{0}})
}{{\rm Li}_{1/2}(-e^{\beta \mu })}\tilde{\nabla}\tilde{n}_{e}  \nonumber \\
+\frac{H^{2}}{2}\tilde{\nabla}\left[ \frac{1}{\sqrt{\tilde{n}_{e}}}\left( 
\frac{\partial ^{2}}{\partial \tilde{x}^{2}}+\frac{\partial ^{2}}{\partial 
\tilde{y}^{2}}\right) \sqrt{\tilde{n}_{e}}\right] \,,  \label{e34}
\\
\left( \frac{\partial ^{2}}{\partial \tilde{x}^{2}}+\frac{\partial ^{2}}{
\partial \tilde{y}^{2}}\right) \tilde{\phi}=\tilde{n}_{e}-\tilde{n}_{i}\,,
\label{e35}
\end{eqnarray}
where $\Omega =\omega _{ci}/\omega _{pi}$ has been defined and where $\tilde{\nabla}=(\partial /\partial \tilde{x},\partial /\partial \tilde{y},0)$, while Eq. (\ref{e121}) becomes 
\begin{equation}
\tilde{n}_{e}=\frac{{\rm Li}_{3/2}(-e^{\beta \mu })}{{\rm Li}_{3/2}(-e^{\beta \mu _{0}})} \,.  \label{e36}
\end{equation}
In the following calculations, for brevity the tilde sign used for defining
normalized quantities will be omitted.

In order to find a nonlinear evolution equation describing the magnetized
plasma, the stretching of the independent variables $x,$ $y$ and $t$ is
defined under the assumption of strong magnetization 
as follows \cite{r34,Mace,r36,r361},
\begin{equation}
X=\varepsilon ^{1/2}(x-V_{0}t) \,, \quad Y=\varepsilon ^{1/2}y \,, \quad \tau =\varepsilon ^{3/2}t \,,
\end{equation}
where $\varepsilon $ is a formal small expansion parameter and $V_{0}$ is
the phase velocity of the wave, to be determined later on. The perturbed
quantities can be expanded in powers of $\varepsilon $ as follows, 
\begin{eqnarray}
n_{j}=1+\varepsilon n_{j1}+\varepsilon ^{2}n_{j2}+...\,, \quad j=e,i \,, 
\\
u_{ix}=\varepsilon u_{x1}+\varepsilon ^{2}u_{x2}+\varepsilon ^{3}u_{x3}...\,, 
\\
u_{i\perp }=\varepsilon ^{3/2}u_{\perp 1}+\varepsilon ^{2}u_{\perp
2}+\varepsilon ^{5/2}u_{\perp 3}...\,, \quad \perp =y,z \,, 
\\
\phi =\varepsilon \phi _{1}+\varepsilon ^{2}\phi _{2}+...\,, 
\\
\mu =\mu _{0}+\varepsilon \mu _{1}+\varepsilon ^{2}\mu _{2}+...  \label{e37}
\end{eqnarray}
In the present model, the ion velocity components ($u_{iy}$, $u_{iz}$) in
the perpendicular to the magnetic field directions are taken as higher order
perturbations compared to the parallel component $u_{ix}$ since in the
presence of a strong magnetic field, the plasma is anisotropic so that the
ion gyro-motion becomes a higher order effect.

The lowest $\varepsilon $ order terms $(\sim \varepsilon ^{3/2})$ from the
set of equations (\ref{e30})-(\ref{e34}) give 
\begin{eqnarray}
-V_{0}\frac{\partial n_{i1}}{\partial X}+\frac{\partial u_{x1}}{\partial X}=0 \,,
\label{e371}
\\
V_{0}\frac{\partial u_{x1}}{\partial X}=\frac{\partial \phi _{1}}{\partial X} \,,
\label{e372}
\\
u_{z1}=\frac{1}{\Omega }\frac{\partial \phi _{1}}{\partial Y}  \,, 
\label{e373}
\\
-\Omega u_{y1}=0  \,,
\label{e374}
\\
\frac{\partial \phi _{1}}{\partial X}=\frac{\partial n_{e1}}{\partial X} \,.
\label{e375}
\end{eqnarray}
The velocity $u_{z1}$  appears in Eq. (\ref{e373}) due to the $E\times B$ drift. 

The lowest $\varepsilon $ order terms $(\sim \varepsilon )$ from equations (\ref{e35}) and (\ref{e36}) give 
\begin{equation}
n_{i1} = n_{e1} = \frac{{\rm Li}_{1/2}(-e^{\beta \mu _{0}})}{{\rm Li}_{3/2}(-e^{\beta \mu _{0}})}\,\beta\mu_1 \,.  \label{e376}
\end{equation}
Solving the system (\ref{e371})-(\ref{e376}), we get $V_{0}=\pm 1$. We set $V_{0}=1$ (the normalized phase velocity of the
MIAW) without loss of generality.

Collecting the next higher order terms of the ion continuity ($\sim
\varepsilon ^{5/2}$) and of the X , Y and Z components of the ion momentum equations ($\sim 
\varepsilon ^{5/2},\varepsilon ^{2},\varepsilon ^{2}$), and after a
rearrangement, we find
\begin{eqnarray}
\frac{\partial n_{i1}}{\partial \tau }-\frac{\partial n_{i2}}{\partial X}+\frac{\partial u_{x2}}{\partial X}+\frac{\partial }{\partial X}
(n_{i1}u_{x1})+\frac{\partial u_{y2}}{\partial Y}=0 \,, \label{e38}
\\
\frac{\partial u_{x1}}{\partial \tau }-\frac{\partial u_{x2}}{\partial X
}+u_{x1}\frac{\partial }{\partial X}u_{x1}=-\frac{\partial \phi _{2}}{\partial X}  \,, \label{e381}
\\
-\frac{\partial u_{y1}}{\partial X}=\Omega u_{z2}  \,, \label{e382}
\\
\frac{\partial u_{z1}}{\partial X}=\Omega u_{y2}  \,. \label{e383}
\end{eqnarray}
Using Eq. (\ref{e376}) and the next higher order terms $\sim \varepsilon ^{5/2}$ from the equations of motion
of the inertialess degenerate electrons in the X and Y directions, we get
\begin{equation}
\frac{\partial n_{e2}}{\partial X}=\frac{\partial \phi _{2}}{\partial X}
+\alpha n_{e1}\frac{\partial n_{e1}}{\partial X}+\frac{H^{2}}{4}\left( \frac{
\partial ^{3}}{\partial X^{3}}+\frac{\partial }{\partial X}\frac{\partial
^{2}}{\partial Y^{2}}\right) n_{e1}  \label{e39}
\end{equation}
and 
\begin{equation}
\frac{\partial n_{e2}}{\partial Y}=\frac{\partial \phi _{2}}{\partial Y}
+\alpha n_{e1}\frac{\partial n_{e1}}{\partial Y}+\frac{H^{2}}{4}\left( \frac{\partial }{\partial Y}\frac{\partial ^{2}}{\partial X^{2}}+\frac{\partial
^{3}}{\partial Y^{3}}\right) n_{e1},  \label{e391}
\end{equation}
where $\alpha $ has been defined in equation (\ref{e61}).

Now collecting the $\varepsilon ^{2}$\ order terms from Poisson's equation,
we have
\begin{equation}
\left( \frac{\partial ^{2}}{\partial X^{2}}+\frac{\partial ^{2}}{\partial
Y^{2}}\right) \phi _{1}=n_{e2}-n_{i2} \,, \label{e392}
\end{equation}
while the next higher terms from Eqs. (\ref{e38}), (\ref{e381}) and 
(\ref{e383}) give
\begin{eqnarray}
\frac{\partial n_{i2}}{\partial X} &=& \frac{\partial n_{i1}}{\partial \tau }+\frac{\partial }{\partial X}(n_{i1}u_{x1})+
\frac{1}{\Omega }\frac{\partial }{\partial Y}\left( \frac{\partial u_{z1}}{\partial X}\right)   \nonumber \\
&&+\frac{\partial u_{x1}}{\partial \tau }+u_{x1}\frac{\partial u_{x1}}{\partial X}+\frac{\partial \phi _{2}}{\partial X} \,.  \label{e393}
\end{eqnarray}
Differentiating Eq. (\ref{e392}) with respect to $X$ and using
Eqs. (\ref{e39}) and (\ref{e393}) together with $n_{i1}=n_{e1}=u_{x1}=\phi _{1}$, $u_{z1}=(1/\Omega)\,\partial \phi
_{1}/\partial Y$, it is finally possible to write the ZK
equation for obliquely propagating quantum MIAWs in terms of $\phi
_{1}\equiv \varphi $, 
\begin{equation}
\frac{\partial \varphi }{\partial \tau }+A\varphi \frac{\partial \varphi }
{\partial X}+\frac{\partial }{\partial X}\left( B\frac{\partial ^{2}\varphi }
{\partial X^{2}}+C\frac{\partial ^{2}\varphi }{\partial Y^{2}}\right) =0 \,.
\label{e40}
\end{equation}
The nonlinearity coefficient $A$ and the dispersion coefficients $B$
and $C$ in the parallel and perpendicular directions of the magnetic field,
respectively, are defined as 
\begin{eqnarray}
A=\frac{1}{2}\left( 3-\alpha \right) \,,  \label{e41}
\\
B=\frac{1}{2}\left( 1-\frac{H^{2}}{4}\right) \,,  \label{e42}
\\
C=\frac{1}{2}\left( 1+\frac{1}{\Omega ^{2}}-\frac{H^{2}}{4}\right) \,.
\label{e43}
\end{eqnarray}
In the purely classical limit the nonlinearity and dispersion coefficients
become $A=1,$ $B=1/2$ and $C=(1/2)\left( 1+1/\Omega ^{2}\right) $ in agreement
with Refs. \cite{r20, r36,r361} treating MIAWs in a classical electron-ion plasma. 
In addition, the ZK equation for fully degenerate plasma will have $A = 4/3$ and $H  = 
(1/2)\,\hbar\omega_{pe}/E_F$ in the coefficients $B, C$. The associated fully degenerate ZK equation
does not matches the results from Refs. \cite{mos1, mos2, kha}, after comparison using physical 
(dimensional and non-stretched) coordinates, restricted to the case of 
electron-ion plasmas. Note that the ZK equations from the previous works do not match the purely classical 
result. 

Provided $l_{x}^{2}B+l_{y}^{2}C\neq 0$, the soliton solution of the ZK
equation (\ref{e40}) for obliquely propagating MIAWs is given by 
\begin{equation}
\varphi =\varphi _{0}\,{\rm sech}^{2}(\eta /W) \,, \label{e45}
\end{equation}
where $\varphi _{0}=3u_{0}/(Al_{x})$ is the height and where $W=
\sqrt{4l_{x}(l_{x}^{2}B+l_{y}^{2}C)/u_{0}}$ is the width of the soliton in terms
of the stretched coordinates. The polarity of the soliton depends on the
sign of $\varphi _{0}$. The transformed coordinate $\eta $ in the co-moving
frame is defined as $\eta =l_{x}X+l_{y}Y-u_{0}\tau $, where $u_{0}\neq 0$ is
the speed of the nonlinear pulse and where $l_{x}>0$ and $l_{y}$ are
direction cosines, so that $l_{x}^{2}+l_{y}^{2}=1$. To obtain localized
structures, decaying boundary conditions 
as $\eta \rightarrow \pm \infty $ were applied. 
Following the habitual usage the terminology ``soliton" is
applied to the solitary wave (\ref{e45}), although the ZK equation does not
belong to the class of completely integrable evolution equations. The dispersion effects arising from the
combination of charge separation and finite ion Larmor radius balances the
nonlinearity in the system to form the soliton.

Defining $\delta V = \varepsilon u_0/l_x$, in the laboratory frame the
solution reads 
\begin{eqnarray}
\varphi = \frac{3\,\delta V}{A}\,{\rm sech}^{2}\Bigl\{\frac{1}{2}\left(\frac{%
\delta V}{l_{x}^2 B + l_{y}^2 C}\right)^{1/2} \times \nonumber \\ \times \Bigl[l_x \Bigl(x - (V_0 +
\delta V)\,t\Bigr) + l_y y\Bigr]\Bigr\} \,.  \label{lab}
\end{eqnarray}
It is apparent that $V_0 + \delta V$ corresponds to the velocity at which
travels the intersection between a plane of constant phase and a field line,
down the same field line \cite{Mace}.

From Eq. (\ref{lab}) the width $L$ of the soliton in the
laboratory frame is 
\begin{equation}  \label{wid}
L = 2 \left(\frac{l_{x}^2 B + l_{y}^2 C}{\delta V}\right)^{1/2} = \sqrt{%
\frac{2}{\delta V}}\left(1-\frac{H^2}{4}+\frac{l_{y}^2}{\Omega^2}%
\right)^{1/2}
\end{equation}
From this expression one concludes that while in the non-quantum $H = 0$
limit necessarily $\delta V > 0$ (bright soliton propagating at supersonic
speed), in the quantum case one has the theoretical possibility of $\delta V < 0$ (dark
soliton propagating at subsonic speed), provided $H^2/4 > 1 +
l_{y}^2/\Omega^2$. 

It is interesting to single out the different quantum effects in the soliton (\ref{lab}). The amplitude/dip of bright/dark solitons is inversely proportional to $A$, which depends only on quantum degeneracy and not on quantum diffraction. More degenerate systems produce a smaller $\alpha$ - as seen e.g. from Fig. 1 - and hence a bigger $A$. Therefore the soliton amplitude/dip decreases with the degeneracy. On the other hand, from the width (\ref{wid}) one find the dependence on the quantum diffraction parameter $H^2$ (which is also dependent on the degeneracy degree) shown in Fig. \ref{fig3} below. For semiclassical bright solitons ($\delta V > 0$) one has $L^2$ decreasing with quantum effects, while for quantum, dark solitons ($\delta V < 0$) one has $L^2$ increasing with quantum effects. 

\begin{figure}[!hbt]
\begin{center}
\includegraphics[width=8.0cm,height=6.0cm]{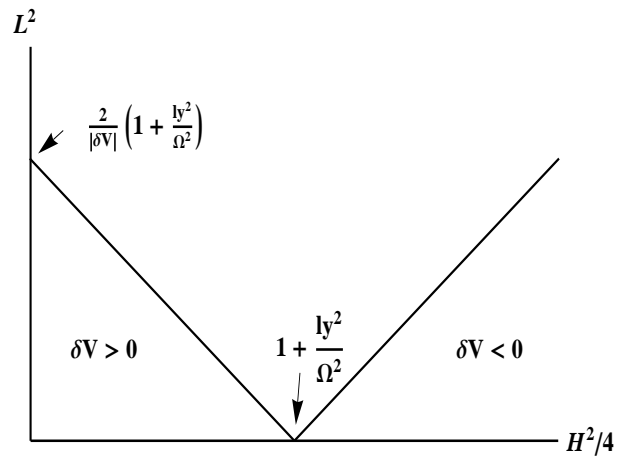}
\end{center}
\caption{Squared width of the localized structure as shown in Eq. (\ref{wid}), as a function of the quantum diffraction parameter $H^2$.}
\label{fig3}
\end{figure}

In unmagnetized quantum plasmas, by construction the term $\sim 1/\Omega^2$
does not appears. In this case when $H = 2$ the corresponding Korteweg-de
Vries (KdV) equation collapses to the Burger's equation, producing an
ion-acoustic shock wave structure instead of a soliton \cite{r8}.

In the magnetized case, a further possibility happens when $C = 0$, or,
equivalently, $1+1/\Omega^2 = H^2/4$, which is not allowed in the classical
limit ($H \equiv 0$). When $C = 0$, from Eq. (\ref{e40}) one has 
\begin{equation}
\frac{\partial \varphi}{\partial \tau }+A\varphi\frac{\partial \varphi}{\partial X} - \frac{1}{2\Omega^2}\frac{\partial^3 \varphi}{\partial X^3} = 0
\,,
\label{kd}
\end{equation}
which transforms to the KdV equation in its standard form by means of $\varphi \rightarrow - \varphi, X \rightarrow - X, \tau \rightarrow \tau$.
Therefore, in this particular situation the problem becomes completely
integrable.

Finally, if $B=0$, it means that $1-H^{2}/4=0$ due to which 
$C=1/2\Omega ^{2}$, so that Eq. (\ref{e40}) becomes
\begin{equation}
\frac{\partial \varphi }{\partial \tau }+A\varphi \frac{\partial \varphi }{%
\partial X}+\frac{1}{2\Omega ^{2}}\frac{\partial }{\partial X}\frac{\partial
^{2}\varphi }{\partial Y^{2}}=0,
\end{equation}
This is a KdV-like equation having perpendicular to the magnetic field dispersion effects, due to the obliquely
propagating MIAW.

Regarding the ranges of validity of the parameters $A, B$ and $C$ in Eqs. (\ref{e41}--\ref{e43}), first we observe that from Eq. (\ref{e61}) one has $1/3 < \alpha < 1$, so that $1 < A < 4/3$. In addition, since $H^2$ from Eq. (\ref{hh}) can in principle attain any non-negative value, $B$ and $C$ are not positive definite. However, strictly speaking, very large values of $H^2$ are associated with strongly coupling effects which can have a large impact on soliton propagation, to be addressed in a separate extended theory. Only in such generalized framework one could be able to make more precise statements on pure quantum soliton existence or non-existence. Nevertheless, significant values of $H^2$ are certainly physically acceptable, as found from the present treatment. These issues are best discussed in the following Section.

\section{Applications}
\label{app}

It is important to discuss the validity domain of the general theory devised in the last Sections. Moreover, it is highly desirable to offer precise physical parameters where the predicted linear and nonlinear waves could be searched in practice. Obviously, the theory is more relevant in the intermediate regimes, where the thermal and Fermi temperatures are not significantly different. Otherwise, the fully degenerate or dilute limits could be sufficiently accurate. Therefore, in this Section frequently we assume 
\begin{equation}
T = T_F \,,
\end{equation}
where $T_F = E_F/\kappa_B$ is the electrons Fermi temperature. 

To start, consider the normalization condition (\ref{muu}), which can be expressed \cite{r33} as 
\begin{equation}
\label{n}
{\rm Li}_{3/2}(-z) = - \frac{4}{3\sqrt{\pi}}\,(\beta E_F)^{3/2} \,,
\end{equation}
where $z = \exp(\beta\mu_0)$. 
For equal thermal and Fermi temperatures, $\beta E_F = 1$, which from Eq. (\ref{n}) gives the equilibrium fugacity $z = 0.98$ and from 
Eq. (\ref{e61}) a parameter $\alpha = 0.80$, 
definitely in the intermediate dilute-degenerate situation as explicitly seen e.g. in Fig. \ref{fig1}.  

Besides quantum degeneracy, quantum diffraction effects can also provide qualitatively new aspects as found e.g. in the extra dispersion of linear waves in Section III and the modified width of the solitons in Section \ref{zkd}. Therefore it would be interesting to investigate systems with a large parameter $H$. However, realistically speaking it is not possible to increase $H$ without limits, which would enter the strongly coupled plasma regime, not included in the present formalism. For instance, the ideal Fermi gas equation of state for electrons would be  unappropriated. Therefore, it is necessary to analyze the coupling parameter $g$ for electrons, which can be defined \cite{Kremp} as 
$g = l/a$, where 
\begin{equation}
\label{ll}
l = - \frac{e^2}{12\pi\varepsilon_0\kappa_B T}\,\frac{n_0 \Lambda_{T}^3}{{\rm Li}_{5/2}(-z)}
\end{equation}
is a generalized Landau length involving the thermal de Broglie wavelength $\Lambda_T = [2\pi\hbar^2/(m_e\kappa_B T)]^{1/2}$, and $a = (4\pi n_0/3)^{-1/3}$ is the Wigner-Seitz radius. 
In the dilute case, one has $e^2/(4\pi\varepsilon_0\,l) =(3/2)\kappa_B T$, so that $l$ would be the classical distance of closest approach in a binary collision, for average kinetic energy. The general expression (\ref{ll}) accounts for the degeneracy effects on the mean kinetic energy. A few calculations yield 
\begin{equation}
\label{gg}
g = - \frac{2\alpha_F\,\sqrt{2 \beta m_e c^2}}{3\,(3\sqrt{\pi})^{1/3}}\,\,\frac{[{\rm Li}_{3/2}^{2}(-z)]^{2/3}}{{\rm Li}_{5/2}(-z)} \,,
\end{equation}
an expression similar to the one for $H^2$ in Eq. (\ref{hh}). Hence, it is legitimate to suspect that the indiscriminate increase of quantum diffraction gives rise to nonideality effects such as dynamical screening and bound states \cite{Kremp}. Incidentally Eq. (\ref{gg}) agrees with Eq. (16) of \cite{pre}, found from related but different methods. 

The resemblance between the coupling and quantum diffraction parameters is confirmed in Fig. \ref{fig4} below, which can be compared to Fig. \ref{fig2} for $H^2$. Moreover, it is apparent in Fig. \ref{fig5}, that we have $H^2 < 0.5$ for the whole span of degeneracy regimes, as far as $g < 1$. Strictly speaking, the dark soliton, shock wave and the completely integrable case associated to the KdV equation (\ref{kd}) are therefore outside the validity domain of the model, since they need large values of quantum diffraction parameter $H$. Nevertheless, the influence of the wave nature of the electrons can still provide important corrections by its own, at least for reasonable values of $H^2$, as is obvious, for instance, in the width of the ZK soliton in Eq. (\ref{wid}). 

\begin{figure}[!hbt]
\begin{center}
\includegraphics[width=8.0cm,height=6.0cm]{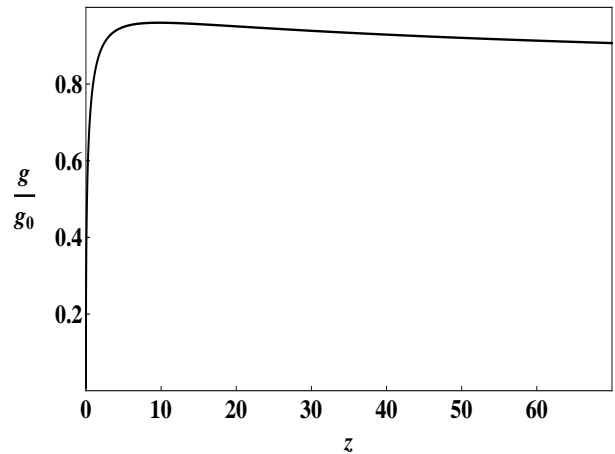}
\end{center}
\caption{Coupling parameter defined in Eq. (\ref{gg}), as
a function of the equilibrium fugacity $z = e^{\beta\mu_0}$, and
normalized to $g_0 = 2 \alpha_F \sqrt{2\beta m_e c^2}/[3(3\sqrt{\pi})^{1/3}]$.}
\label{fig4}
\end{figure}

\begin{figure}[!hbt]
\begin{center}
\includegraphics[width=8.0cm,height=6.0cm]{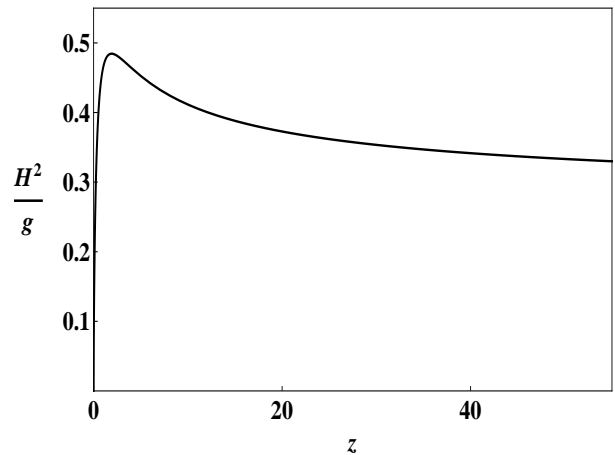}
\end{center}
\caption{Ratio between the diffraction parameter $H^2$ from Eq. (\ref{hh}) and the coupling parameter $g$ from Eq. (\ref{gg}), as
a function of the equilibrium fugacity $z = e^{\beta\mu_0}$.}
\label{fig5}
\end{figure}

\begin{figure}[!hbt]
\begin{center}
\includegraphics[width=8.0cm,height=6.0cm]{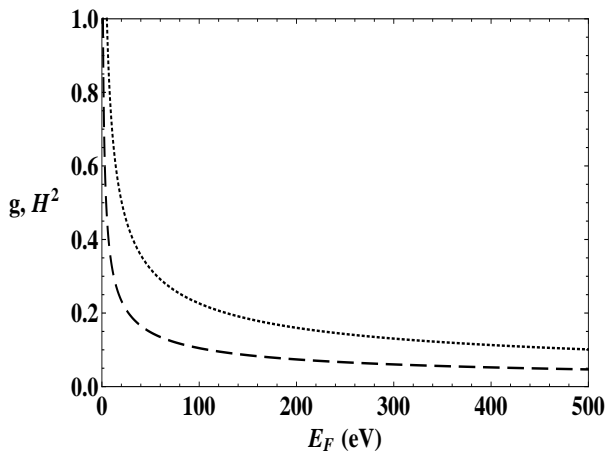}
\end{center}
\caption{Upper, dotted curve: coupling parameter $g$ from Eq. (\ref{gg}); lower, dashed curve: quantum diffraction parameter $H^2$ from 
Eq. (\ref{hh}), as a function of the electronic Fermi energy $E_F$ in ${\rm eV}$, for hydrogen plasma and the intermediate dilute-degenerate regime where $T = T_F$.}
\label{fig6}
\end{figure}

The behavior of the parameters $g, H^2$ can be summarized in Fig. \ref{fig6}, for $T = T_F$ and considering hydrogen plasma parameters. We observe that $g < 1$ for $n_0 > 5.23 \times 10^{28} \,{\rm m}^{-3}$, or $E_F > 5.11 \, {\rm eV}$, which starts becoming realizable for typical densities in solid-density plasmas \cite{Tata, Wagner}. 

It is necessary to discuss additional points about the validity
conditions of the model. Both the static electronic response and long
wavelength (and hence fluid) assumptions are collected in Eq.
(27) of \cite{pre}, reproduced here for convenience, 
\begin{equation}  \label{val}
k_{\rm min} \equiv \frac{2 \sqrt{3} m_{e} c_{s}}{\hbar} \ll k \ll \frac{\omega_{pi}}{c_{s}} \equiv k_{\rm max} 
\,.
\end{equation}
For hydrogen plasma and $T = T_F$, from Eq. (\ref{val}) one has $k_{\rm max} > k_{\rm min}$ for $n_0 < 9.81 \times 10^{35} \, {\rm m}^{-3}$. The later condition is safely satisfied for non-relativistic plasma. At such
high densities available in compact astrophysical objects like white dwarfs
and neutron stars, the Fermi momentum becomes comparable to $m_e c$, asking for a relativistic treatment. Strictly speaking, the first inequality in Eq. (\ref{val}) could be removed, but in this case the quantum recoil would be less significant. In this case, only quantum degeneracy effects would be relevant.

\begin{figure}[!hbt]
\begin{center}
\includegraphics[width=8.0cm,height=6.0cm]{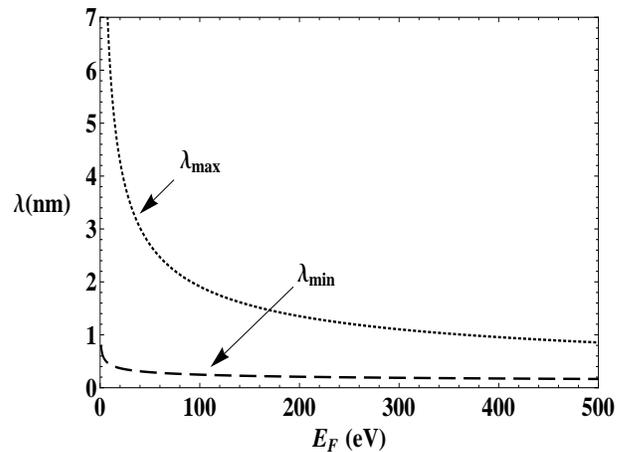}
\end{center}
\caption{Upper, dotted curve: maximum wavelength $\lambda_{\rm max}$; lower, dashed curve: minimal wavelength $\lambda_{\rm min}$, consistent with Eq. (\ref{val}), as a function of the electronic Fermi energy $E_F$ in ${\rm eV}$, for hydrogen plasma and the intermediate dilute-degenerate regime where $T = T_F$.}
\label{fig7}
\end{figure}

In terms of the wavelength $\lambda$, Eq. (\ref{val}) yields a suitable range $\lambda_{\rm min} = 2\pi/k_{\rm max} \ll \lambda \ll \lambda_{\rm max} = 2\pi/k_{\rm min}$, shown in Fig. 
\ref{fig7}, in the nanometric scale from extreme ultraviolet to soft X-rays. 

As discussed in
Section \ref{ine}, the model also assumes that the pressure effects are
significantly larger than magnetic field effects regarding electrons, or 
$k^2 v_{T}^{2}(k) \gg \omega_{ce}^2$. In the worse case where $k \approx k_{\rm min}$, the quantum diffraction is typically a correction in the expression (\ref{vt}) for $v_{T}(k)$. We then find
\begin{equation}
\label{cond}
\frac{\hbar\omega_{ce}}{E_F} \ll 2\sqrt{3}\,\left(\frac{m_e}{m_i}\right)^{1/2}\frac{{\rm Li}_{3/2}(-z)}{\rm Li_{1/2}(-z)} \approx 0.10 \,,
\end{equation}
for $T = T_F$. The condition (\ref{cond}), which is easier to satisfy for larger densities, is safely satisfied except for very strong magnetic fields. For instance, for $E_F \approx 10 \, {\rm eV}$, we just need $B_0 \ll 8.81 \, {\rm kT}$. 

Ions have been assumed to be cold, classical and ideal (disregarding strong ion coupling effects). Denoting $T_i$ as the ionic temperature, when $T_F \gg T_i$ the MIAW phase speed becomes much larger than the ionic thermal speed, justifying the cold ions assumption. On the other hand, there is the need of a small ionic coupling parameter $g_i$. For a non-degenerate ionic fluid we then \cite{Kremp} have  
\begin{equation}
g_i = \frac{e^2}{4\pi\varepsilon_0 a}\,\left(\frac{3\kappa_B T_i}{2}\right)^{-1} \ll 1 \,.
\end{equation}
Otherwise one would have a ionic liquid or even an ionic crystal, as is believed to happens for $g_i \approx 172$ in an one-component plasma \cite{Murillo}. The joint requirements of cold and non-strongly coupled ions is represented in Fig. \ref{fig8} below, where the allowable region is between the straight lines $T_i = T_F, g_i = 1$. 
A minimal number density $n_0 = 3.63 \times 10^{28}\,{\rm m}^{-3}$ is found to be necessary, which again is accessible for typical solid-density plasmas. 

\begin{figure}[!hbt]
\begin{center}
\includegraphics[width=8.0cm,height=6.0cm]{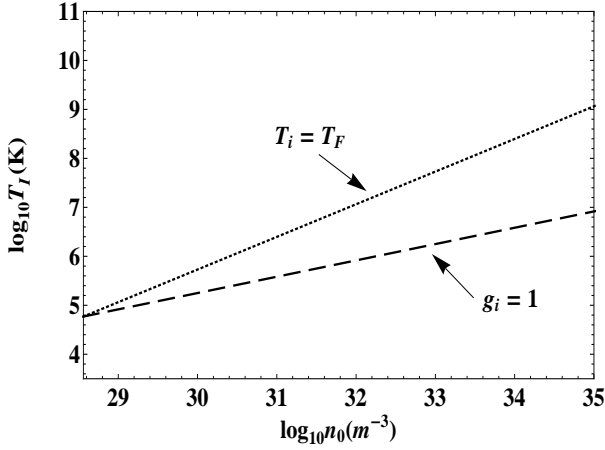}
\end{center}
\caption{Cold and weakly coupled ions are below the upper, dotted straight line (where the ionic temperature $T_i$ equals the electronic Fermi temperature $T_F$) and above the lower, dashed straight line (where the ionic coupling parameter $g_i = 1$).}
\label{fig8}
\end{figure}

For the sake of illustration, consider the bright soliton solution from Eq. (\ref{lab}), for parameters representative of solid-density plasmas, namely, $n_0 = 5 \times 10^{30} \, {\rm m}^{-3}, H^2 = 0.10, g = 0.22$ and $T = T_F = 1.24 \times 10^{6} \, {\rm K}$. In this case the allowable wavenumbers satisfy $0.24 \, {\rm nm} < \lambda < 1.85 \, {\rm nm}$, the ionic temperatures are in the range $3.04 \times 10^5 \, {\rm K} < T_i < 1.57 \times 10^6 \, {\rm K}$ and the magnetic field should have a strength $B_0 < 9.23 \times 10^{4} \, {\rm T}$. 

\section{Summary}

The main results of the work are the dispersion relation (\ref{w}) and the
ZK equation (\ref{e40}), both of which describe quantum MIAWs in the linear
and nonlinear regimes, respectively, allowing for arbitrary electrons
degeneracy degree. The results significantly generalize the previous
literature, restricted to either dilute (Maxwell-Boltzmann) or fully
degenerate plasmas. The conditions for applications are investigated in depth, 
pointing the importance of the findings, for e.g. magnetized solid-density plasma. 
While the physical parameters were more focused on hydrogen plasma in the 
intermediate dilute-degenerate limit where $T = T_F$, adaptation to fully ionized electron-ion 
cases with atomic number $Z \neq 1$ and arbitrary temperatures is not difficult at all.  
With minor remarks, the whole parametric analysis of Section \ref{app} also applies to 
the unmagnetized case. It is hoped that the detailed assessment of physical parameters 
thus developed, will incentive experimental and observational verifications of linear and 
nonlinear quantum ion-acoustic waves, 
both in laboratory and nature, considering a large span of degeneracy regimes. 
A possibly important next step is the rigorous (non ad hoc) incorporation of
exchange-correlation effects, which are beyond the reach of the present
communication.

\acknowledgments

FH acknowledges CNPq (Conselho Nacional de Desenvolvimento Cient\'{\i}fico e
Tecnol\'{o}gico) for financial support.

\appendix

\section{Derivation of Eq. (\ref{sum})}

From Eqs. (\ref{kce}) and (\ref{fd}) one has 
\begin{eqnarray}
&\strut& \chi_{e}(0,{\bf k})  \nonumber \\
&=& \frac{{\cal A} e^2}{\varepsilon_0 \hbar k^2}\!\!\int\!\! d{\bf v}\! \left[\frac{1}{{\bf k}\cdot{\bf v} + \hbar k^2/(2 m_e)} - \frac{1}{{\bf k}
\cdot{\bf v} - \hbar k^2/(2 m_e)}\right]\times  \nonumber \\
&\times& \frac{1}{1 + \exp(\beta m_e v^2/2)/z}  \nonumber \\
&=& \!\frac{2\pi {\cal A} e^2}{\varepsilon_0 \hbar k^2}\!\!\int_{-\infty}^{\infty}
\!\!\!\!\!dv_\parallel\!\!\left[\frac{1}{k v_\parallel + \hbar k^2/(2 m_e)}
- \frac{1}{k v_\parallel - \hbar k^2/(2 m_e)}\right]\times  \nonumber \\
&\times& \int_{0}^{\infty} dv_\perp \,\frac{v_\perp}{1 + \exp[\beta m_e
(v_{\perp}^2+v_{\parallel}^2)/2]/z} \,,
\end{eqnarray}
where ${\bf v} = v_{\parallel} {\bf k}/k + {\bf v}_\perp \,, 
{\bf k}\cdot{\bf v}_\perp = 0$, $z = \exp(\beta\mu_0)$ and all
integrals consider the principal value sense.

Performing the $v_\perp$ integral, considering the expression of ${\cal A}$ in Eq. (\ref{muu}) and 
applying a simple change of variables we get 
\begin{eqnarray}
\chi_{e}(0,{\bf k}) &=& \frac{\beta m_e \omega_{pe}^2}{4 \sqrt{\pi} \,
{\rm Li}_{3/2}(-z) q k^2} \int_{-\infty}^{\infty} \frac{ds}{s} \times
\label{a2} \\
&\times& \left[\ln\left(1 + z e^{-(s+q)^2}\right) - \ln\left(1 + z
e^{-(s-q)^2}\right)\right] \,,  \nonumber
\end{eqnarray}
where $q = \sqrt{\beta/(2 m_e)} \, \hbar k/2$.

Expanding in powers of the quantum recoil one has 
\begin{eqnarray}
\ln\Bigl(1 &+& z e^{-(s+q)^2}\Bigr) - \ln\Bigl(1 + z e^{-(s-q)^2}\Bigr)
\label{log} \\
&=& - 4 q \Bigl(g(s) + \frac{q^2}{3!}g^{\prime \prime }(s) + \frac{q^4}{5!}
g^{(iv)}(s) + \mathcal{O}(q^6)\Bigr) \,,  \nonumber
\end{eqnarray}
where $g(s) \equiv s/[1 + \exp(s^2)/z]$. At this point notice that the
possible divergence at $s = 0$ in the integral (\ref{a2}) was explicitly
removed.

After integrating by parts, it is found that 
\begin{eqnarray}
\chi_{e}(0,{\bf k}) &=& - \frac{2\beta m_e \omega_{pe}^2 z}{\sqrt{\pi} \,
{\rm Li}_{3/2}(-z) k^2} \Bigl(\int_{0}^{\infty} \frac{ds}{z + e^{s^2}} 
\nonumber \\
&-& \frac{2 q^2}{3}\int_{0}^{\infty} \frac{ds \,e^{s^2}}{(z + e^{s^2})^2}
\label{fk} \\
&+& \frac{4 q^4}{15}\int_{0}^{\infty} \frac{ds \,e^{s^2} (e^{s^2}-z)}{(z +
e^{s^2})^3} + \mathcal{O}(q^6) \Bigr) \,.  \nonumber
\end{eqnarray}
From Eqs. (\ref{e9}) and (\ref{shi}), each term on the right hand side of (\ref{fk}) can be evaluated in terms of polylogarithms. For instance, 
\begin{eqnarray}
{\rm Li}_{1/2}(-z) &=& - \frac{2 z}{\sqrt{\pi}}\int_{0}^{\infty}\frac{ds}{z+e^{s^2}} \,,  \nonumber \\
{\rm Li}_{-1/2}(-z) &=& - \frac{2 z}{\sqrt{\pi}}\int_{0}^{\infty}\frac{ds\, e^{s^2}}{(z+e^{s^2})^2} \,,  \\
{\rm Li}_{-3/2}(-z) &=& \frac{2 z}{\sqrt{\pi}}\int_{0}^{\infty}\frac{ds\,
e^{s^2} (z - e^{s^2})}{(z+e^{s^2})^3} \,,  \nonumber
\end{eqnarray}
clearly related to the integrals in expression (\ref{fk}). In this
way Eq. (\ref{sum}), which is also independently confirmed by numerical evaluation, is proved.

\end{document}